\pdfoutput=1 
\RequirePackage{build-config}
\newcommand{\formattype}{\formattypeIEEE}

\RequirePackage{ifthen}

\newcommand{\formattypeIEEE}{formattypeIEEE}

\ifisdraft
  \ifthenelse{\equal{\formattype}{\formattypeIEEE}}{
    \documentclass[10pt,conference]{IEEEtran}
  } {
    \ifisarxiv
      \ifisanonymous
        \documentclass[acmsmall,screen,nonacm,review,anonymous]{acmart/acmart}
        \settopmatter{printfolios=true,printccs=false,printacmref=false}
      \else
        \documentclass[acmsmall,screen,nonacm,review]{acmart/acmart}
        \settopmatter{printfolios=true,printccs=false,printacmref=false}
      \fi
    \else
      \ifisanonymous
        \documentclass[acmsmall,screen,review,anonymous]{acmart/acmart}
        \settopmatter{printfolios=true,printccs=true,printacmref=true}
      \else
        \documentclass[acmsmall,screen,review]{acmart/acmart}
        \settopmatter{printfolios=true,printccs=true,printacmref=true}
      \fi
    \fi
  }
\else
  \ifthenelse{\equal{\formattype}{\formattypeIEEE}}{
    \documentclass[10pt,conference]{IEEEtran}
  } {
    \ifisarxiv
      \ifisanonymous
        \documentclass[acmsmall,screen,nonacm,anonymous]{acmart/acmart}
        \settopmatter{printfolios=true,printccs=false,printacmref=false}
      \else
        \documentclass[acmsmall,screen,nonacm]{acmart/acmart}
        \settopmatter{printfolios=true,printccs=false,printacmref=false}
      \fi
    \else
      \ifisanonymous
        \documentclass[acmsmall,screen,anonymous]{acmart/acmart}
        \settopmatter{printfolios=true,printccs=true,printacmref=true}
      \else
        \documentclass[acmsmall,screen]{acmart/acmart}
        \settopmatter{printfolios=true,printccs=true,printacmref=true}
      \fi
    \fi
  }
\fi

\ifthenelse{\equal{\formattype}{\formattypeIEEE}}{

} {
  \citestyle{acmauthoryear}
}

\newwrite\abstractoutput
\immediate\openout\abstractoutput=\jobname.abstract.output

\ifthenelse{\equal{\formattype}{\formattypeIEEE}}{
  \newcommand{\makeTitleAndAbstract}{
    \write\abstractoutput{\theAbstract}
    \maketitle
    \begin{abstract}
    \theAbstract
    \end{abstract}

    \begin{IEEEkeywords}
      NPM, dependency-management, JavaScript
    \end{IEEEkeywords}
  }
}{
  \newcommand{\makeTitleAndAbstract}{
    \write\abstractoutput{\theAbstract}
    \begin{abstract}
    \theAbstract
    \end{abstract}
    \maketitle
  }
}

\usepackage{paper-specific-macros}
\usepackage{package-config}
\usepackage{tikz}
\usetikzlibrary{positioning,fit,arrows.meta,backgrounds,shapes.geometric}

\IEEEoverridecommandlockouts

\usepackage{cite} 

\begin{document}

\makeTitleAndAbstract

\section{Introduction}
\label{sec:intro}


Modern software development relies inextricably on open source package repositories on a massive scale.
For example, the NPM repository contains over two million packages and serves tens of billions of downloads weekly,
and practically every JavaScript application uses the NPM package manager to install
packages from the NPM repository.
As open source package repositories grow in scale, the maintenance, updating, and distribution of packages
represents a growing attack surface for malicious actors to target, and understanding the properties
of the software supply chain is vital.

One particular concern in open source ecosystems is the \emph{technical lag}~\cite{msr-resp-c1,msr-resp-c2,msr-resp-b4,msr-resp-b5,technical-lag} that packages experience between when a new update is available for a dependency and when that update is applied.
NPM and other similarly-designed ecosystems (PyPi, etc.) offer a potential solution
in the form of semantic versioning (``semver'') and flexible version constraints.
In semver, versions are numbered in the form \texttt{major.minor.bug}, where \texttt{major} denotes breaking API changes,
\texttt{minor} denotes a non-breaking change adding new functionality, and \texttt{bug} denotes a backwards-compatible bug fix\footnote{In this paper we use ``bug'' rather than the standard ``patch'' semver terminology, so as to disambiguate from the notion of \emph{security patches}.}\cite{semver-spec}.
Flexible version constraints allow developers of downstream (i.e., dependent) packages to specify which types of updates
they are willing to automatically accept.
Ideally, semver helps developers to express constraints and version numbers so that
non-breaking important updates (such as security patches)
flow rapidly to downstream packages, while breaking changes are delayed until
developers choose to accept them.
For example, a developer may specify that they depend on the package \texttt{react}, with constraint \texttt{\^{}18.1.1},
which means that automatic updates are allowed until (excluding) version \texttt{19.0.0}. In essence, this constraint
says ``receive all updates to React that are unlikely to be breaking changes''.

However, there are three significant complications with semver in practice that can lead to technical lag~\cite{msr-resp-c1,msr-resp-c2,msr-resp-b4,msr-resp-b5,technical-lag}.
First, the positive properties of semver are predicated on both upstream developers labeling their updates
with the correct semver increment type, and on downstream developers using constraints that are neither too flexible nor too strict.
Second, dependencies in the middle of a transitive dependency chain affect the final received versions
of dependencies. The downstream developer may list a constraint that allows the most up-to-date version of a package,
but if a transitive dependency has a more restrictive constraint,
the downstream developer may not receive the up-to-date version.
Third, allowing for automatic (bug) updates to dependencies can be dangerous, as it introduces
an attack vector for malware.

In this work, we aim to understand how developers make use of
dependencies, semantic versioning, and flexible version constraints at the ecosystem-scale,
and how all these factors intersect to affect developer experience and
supply chain security.
Prior work on mining data from the NPM ecosystem has primarily focused on answering questions
about NPM at a snapshot in time \cite{weak-links-npm,micro-packages,trivial-packages,trivial-packages-why}.
In this work, we first understand how developers make use of semantic versioning by analyzing flexible constraint type frequency
and semver increment type frequency over the entire history of NPM.
Then, to understand how updates flow in practice at the ecosystem scale,
we run large-scale experiments that resolve packages' dependencies at different snapshots in time, observing how long it takes
for updates to be received by downstream packages.
To enable these experiments, we built a tool that allows for accurate time-travel dependency solving throughout the history of NPM.
This methodology allows for more precision in resolving dependencies throughout time, as prior work~\cite{msr-resp-b5,msr-resp-d1,msr-resp-c2,msr-resp-d2,msr-resp-b3} approximated NPM's behavioral semantics, which are not well-specified~\cite{maxnpm}.

In total, we have built the first dataset of NPM that includes (as of October 31, 2022):
\begin{enumerate}
  \item every package on NPM (2,663,681 packages)
  \item every version of every package (28,941,927 versions)
  \item metadata ($\approx$ 40 GB compressed) and packaged code ($\approx$ 19 TB compressed) for every version of every package,
  \item full data of security advisories issued for NPM packages, downloaded from the GitHub Security Advisory database.
\end{enumerate}
This dataset is indexed to allow for easy querying and large-scale distributed
computations.
To gather this data, we designed and implemented a distributed system for downloading, archiving and retrieving packages from NPM.
We release our scraper and dataset under the BSD 3-Clause license\footnote{
  Please see \url{https://dependencies.science} for access to up-to-date metadata, tarball data, and source code.
  The original artifact excluding tarball data is available on Zenodo~\cite{citeTheArtifact}.
}.

We use our dataset to answer several questions about the NPM ecosystem, in particular how
developers use semantic versioning, and how this affects supply chain security:
\begin{itemize}
  \item \textbf{RQ1}: Do developers specify dependency version constraints to allow for automated updates?
  \item \textbf{RQ2}: Do developers use semantic versioning in their package
        releases to allow for automated updates to downstream packages?
  \item \textbf{RQ3}: Do packages frequently contain out-of-date dependencies?
        And when updates are published, how long until those updates
        are received by downstream packages?
  \item \textbf{RQ4}:
        Among the types of semver updates, what types of high-level changes do developers tend to make?
        How often do developers only update dependencies?
\end{itemize}

These results are impactful for both developers and researchers.
We show that, generally, the NPM ecosystem is effective in terms of efficient distribution of non-breaking updates,
but most packages end up with out-of-date dependencies anyways due to the sheer volume
of dependencies and updates to deal with.
In addition, we found evidence that some developers use semver non-optimally when releasing security patches,
and that minor and major semver updates appear to have a higher risk of introducing security vulnerabilities.

\section{Related Work}
\label{sec:related-work}


Our research questions and methodology build on a large body of related work examining semantic versioning and technical lag.



\subsubsection{Semantic Versioning}
While semantic versioning does have a precise syntactic specification~\cite{semver-spec},
the semantics of what counts as backwards-compatible are not formally defined.
Tooling, including NPM, generally does not enforce how developers make use of semantic versioning in practice.
Choices of semantic versioning usage impact speed
of distribution of packages, technical lag, stability, developer frustration, and more. Developer interviews in 2015 conducted by
Bogart et al.~\cite{stability-dependencies} in the NPM and CRAN ecosystems found that developers try to use semantic versioning,
but are not always aware of its implications and generally find dependency management exhausting.
More concretely, Raemaekers et al. \cite{maven-semver-not-followed}~\cite{maven-semver-not-followed} found that in 2006--2011, Maven developers
often introduced binary incompatible changes within supposedly non-breaking semver updates.
Wittern et al.~\cite{dynamics-of-js-ecosystem} studied dependencies
between packages in NPM, and found that the number of dependencies between packages is increasing over time,
and observed the frequencies of version constraint types in 2016.
Dietrich et al.~\cite{msr-resp-b1} then observed how version constraint type frequencies have changed over time,
at the project level.
Examining version constraint evolution at the full-ecosystem level allows for an evaluation based on ``wisdom of the crowds.''
Decan et al.~\cite{msr-resp-b2} perform an analysis of dependency constraints at the ecosystem level for Cargo, NPM, Packagist and Rubygems.
Focusing only on a single ecosystem (NPM), we validate Decan et al's findings, and perform a much deeper analysis of the dataset.
Our study also examines the frequencies of released update types, which enables us to draw important implications about the diffusion of security updates.




\subsubsection{Technical Lag}
\label{sec:related:technical-lag}

Many pieces of prior work attempt to analyze the propagate of updates to downstream packages,
and how out-of-date the dependencies of a project typically are.
Gonzalez-Barahona et al.~\cite{msr-resp-c1} define the measure
of ``technical lag'', which analyzes how far out-of-date a package's dependencies are relative to more recently released versions,
which has since been been further studied in the context of NPM~\cite{msr-resp-b5,technical-lag,msr-resp-c2}.
In addition, the concept of technical
lag is specialized to the analysis of the propagation of security patches or vulnerabilities in further
work~\cite{msr-resp-d1,msr-resp-b4,msr-resp-d2}.

Calculating technical lag is difficult, and prior works have attempted to simulate the dependencies that would have been resolved at different points in time.
Some of these works do not consider transitive dependencies~\cite{msr-resp-b5,msr-resp-d1}, which is concerning as transitive dependencies typically represent the majority of a package's dependencies in NPM.
Others have followed up by considering transitive dependencies~\cite{msr-resp-c2,msr-resp-d2}.
Liu et al.~\cite{msr-resp-b3} introduce \emph{DTResolver}, a custom dependency solving algorithm that more closely matches
the behavior of NPM.
However, the authors' evaluation of DTResolver found that it only matched NPM's behavior when building dependency trees for 90.58\% of 15,673 libraries~\cite{msr-resp-b3}.
Our recent evaluation of NPM's dependency resolution semantics showed a variety of corner cases in which NPM's algorithm will select unexpected versions for dependencies in order to unify versions~\cite{maxnpm}.
Particularly when resolving transitive dependencies, the error introduced by an incorrect approximation of NPM's resolution semantics compounds.
Compared to \emph{all} prior work that we are aware of in studying technical lag in the NPM ecosystem, ours is the \emph{only} study to use NPM itself to resolve historical dependencies.
We make our tools and dataset available to allow others to employ this methodology~\cite{citeTheArtifact}.

\subsubsection{Studies of NPM} Finally, other studies have looked at more specific questions or applications of data analysis from NPM,
such as studying when developers downgrade packages~\cite{npm-downgrades},
analyzing the phenomenon of popular ``micro'' packages in NPM~\cite{micro-packages,trivial-packages,trivial-packages-why},
and developing methods to understand and prevent vulnerabilities or malware in NPM~\cite{weak-links-npm,amalfi,npm-security-threats,msr-resp-a1}.
We will return to discuss how our findings may guide future research applications in \cref{sec:discussion:researchers}.

\section{Methodology}
\label{sec:methodology}




At a high-level, we answer our four core research questions using different aspects of our dataset and analysis
systems. RQ1 and RQ2 are answered purely via analysis of our scraped metadata.
Answering RQ3 is more challenging as it requires reasoning about how dependencies are resolved across time, which we answer by using
our time-traveling dependency resolver in large-scale experiments.
Finally, to answer RQ4 we compute diffs between tarballs
of package versions.

\subsection{RQ1: Version Constraint Usage}
\label{sec:methodology:version-constraint-usage}
Within NPM's rich language for specifying version constraints on dependencies~\cite{package_json,maxnpm},
it is unclear which of the many constraint types developers frequently make use of
and how loose or restrictive those constraints are.

We classify version constraints in the following mutually exclusive categories:
\begin{enumerate}
  \item Exact constraints (\texttt{"=1.2.3"}) accept no versions other than the specifically listed one;
  \item Bug-flexible constraints (\texttt{"\~{}1.2.3"}) accept any updates to the bug semver component, so \texttt{1.2.4}, etc.;
  \item Minor-flexible constraints (\texttt{"\^{}1.2.3"}) accept any updates to the minor semver component, so \texttt{1.3.0}, etc.;
  \item Geq constraints (\texttt{">=1.2.3"}) accept any versions greater than or equal to the specified version;
  \item Any constraints (\texttt{"*"}) accept any versions; and
  \item Other constraints, such as disjunction, conjunction, GitHub URLs, etc.
\end{enumerate}
We then examine frequencies of these constraint categories across NPM, segmented by year
so we can observe how constraint usage has evolved historically.
In addition, one challenge with analyzing data from NPM is that some packages publish a massive
number of versions (React has over 1,000 versions), so aggregating across all versions
may produce results that are biased towards packages with more versions. In RQ1
we select only the most recent version of every package that was uploaded within
each year. This enables us to segment by time while avoiding this bias.

\subsection{RQ2: Semantic Versioning in Updates}
\label{sec:methodology:package-updates}

We now turn to examine how developers increment their semantic version numbers when publishing updates.
We first find all of the package updates
that have occurred in NPM's history, and classify each as
a bug (e.g. \texttt{5.4.8} $\to$ \texttt{5.4.9}),
minor (e.g. \texttt{5.4.8} $\to$ \texttt{5.5.0}), or
major (e.g. \texttt{5.4.8} $\to$ \texttt{6.0.0}) update.

One would expect that updates can trivially
be identified as consecutive versions of the same package. NPM however allows versions to
be published non-chronologically. This feature allows for maintenance of
parallel version branches. For example, consider the following \emph{chronological} order of versions:
\texttt{1.0.0}, then
\texttt{2.0.0}, then
\texttt{1.0.1}, and then
\texttt{2.0.1}.
In this example, the mined updates should consist of:
\texttt{1.0.0} $\to$ \texttt{2.0.0},\;
\texttt{1.0.0} $\to$ \texttt{1.0.1}, and
\texttt{2.0.0} $\to$ \texttt{2.0.1},
as these reflect updates that are most closely based on the source version while being
chronologically and numerically consistent.
We would not include the update \texttt{1.0.1} $\to$ \texttt{2.0.0} because it is not chronologically
consistent, and thus \texttt{2.0.0} is unlikely to be a derivative of \texttt{1.0.1}.

To determine the set of updates, we group versions by the equivalence relation of same major component
and assert that groups are ordered within themselves chronologically. We then have updates between versions
within each group, and between different groups. Continuing the above example,
we have two groups: $\{\texttt{1.0.0},\, \texttt{1.0.1}\}$ and $\{\texttt{2.0.0},\, \texttt{2.0.1}\}$.
From intra-group ordering we obtain
\texttt{1.0.0} $\to$ \texttt{1.0.1} and
\texttt{2.0.0} $\to$ \texttt{2.0.1},
and from the inter-group ordering we obtain
\texttt{1.0.0} $\to$ \texttt{2.0.0}.
We believe this
algorithm reflects well how developers publish updates, and we discuss alternatives in \cref{sec:threats-to-validity}.
When computing these updates, we first filter out all prerelease versions (e.g. \texttt{1.2.3-beta5}),
yielding 1,453,789 packages with at least one update (of 2,869,085 packages).
We then filter out 52,279 packages that do not have consistent intra-group chronological orders.

With all updates and version increment types identified, we examine the distribution of the three update
types across the whole population, and then compare to the subgroups of updates that introduce and patch
vulnerabilities. Updates that patch vulnerabilities are identified directly in the scraped advisory database,
while we identify versions that introduce vulnerabilities as the minimal version containing that vulnerability.
To avoid the bias introduced by some packages
having a large number of updates, our top-level aggregation is among packages rather than updates.
For each package, we identify the proportion of its updates of each type (segmenting by security effect),
and then visualize this percentage across all the packages. This enables us to make conclusions
about how packages and package developers generally handle incrementing semver numbers during updates.
In addition, note that when segmenting by updates that introduce vulnerabilities, we are \emph{not}
attempting to study malware, rather updates that (probably inadvertently) introduce a vulnerability.

\subsection{RQ3: Out-of-Date Dependencies and Update Flows}
\label{sec:methodology:update-distribution}

The properties examined thus far have been local properties of each package, in that
each package has been analyzed individually.
We now wish to answer how out-of-date NPM packages typically are,
and how long it takes updates to flow to downstream packages.
Both of these properties rely on all the packages in the transitive
dependency closure of a downstream package.
However, reasoning precisely about how dependencies are solved is challenging both because NPM's dependency
solving algorithm is complex (\cref{sec:related:technical-lag}), and because we wish to parameterize this over time.

In order to compute solutions accurately and at different points in time, we use vanilla NPM's solver
combined with a proxy that emulates the world state at any given point in history
(described in \cref{sec:time-travel-impl}).
With this key tool, we then perform two experiments: first we solve the dependencies
of the most recent version of every package in NPM and observe how many packages have
out-of-date dependencies; we then explore how updates flow to downstream packages by
solving the dependencies of the downstream package at different points in time until it receives
the update.

\subsection{RQ4: Analyzing Code Changes in Updates}
\label{sec:methodology:code-changes}

After having examined how developers use constraints and version numbers in isolation,
we next align that with a high-level characterization of what updates actually change.
For every identified update, we decompress the packaged code from both versions,
and look for file changes. We then classify changes as modifying dependencies,
code (\texttt{.js, .ts, .jsx, .tsx}), both, or neither. We then examine the distribution of these
types of changes segmented by semver increment type, again normalizing per-package to avoid
biasing towards packages with more updates.

Analyzing at a deeper level is possible with our dataset, but is beyond the scope of this paper.
Note that many packages upload compiled or minified JavaScript code, which makes it difficult
to even look at simple line-by-line diffs. In addition, we could have chosen to count
other file types as code (\texttt{.sh}, etc.), but we chose to focus on JavaScript code.

\section{System Architecture}
\label{sec:system-architecture}

\tikzset{
  module/.style={%
      draw, rounded corners,
      minimum width=#1,
      minimum height=7mm,
      font=\sffamily
    },
  module/.default=2cm,
  >=LaTeX
}
\tikzset{database/.style={cylinder,aspect=0.5,draw,rotate=90,path picture={
          \draw (path picture bounding box.160) to[out=180,in=180] (path picture bounding
          box.20);
          \draw (path picture bounding box.200) to[out=180,in=180] (path picture bounding
          box.340);
        }}}
\makeatletter
\pgfdeclareshape{document}{
  \inheritsavedanchors[from=rectangle] 
  \inheritanchorborder[from=rectangle]
  \inheritanchor[from=rectangle]{center}
  \inheritanchor[from=rectangle]{north}
  \inheritanchor[from=rectangle]{south}
  \inheritanchor[from=rectangle]{west}
  \inheritanchor[from=rectangle]{east}
  \backgroundpath{
    \southwest \pgf@xa=\pgf@x \pgf@ya=\pgf@y
    \northeast \pgf@xb=\pgf@x \pgf@yb=\pgf@y
    \pgf@xc=\pgf@xb \advance\pgf@xc by-7.5pt 
    \pgf@yc=\pgf@yb \advance\pgf@yc by-7.5pt
    \pgfpathmoveto{\pgfpoint{\pgf@xa}{\pgf@ya}}
    \pgfpathlineto{\pgfpoint{\pgf@xa}{\pgf@yb}}
    \pgfpathlineto{\pgfpoint{\pgf@xc}{\pgf@yb}}
    \pgfpathlineto{\pgfpoint{\pgf@xb}{\pgf@yc}}
    \pgfpathlineto{\pgfpoint{\pgf@xb}{\pgf@ya}}
    \pgfpathclose
    \pgfpathmoveto{\pgfpoint{\pgf@xc}{\pgf@yb}}
    \pgfpathlineto{\pgfpoint{\pgf@xc}{\pgf@yc}}
    \pgfpathlineto{\pgfpoint{\pgf@xb}{\pgf@yc}}
    \pgfpathlineto{\pgfpoint{\pgf@xc}{\pgf@yc}}
  }
}
\makeatother

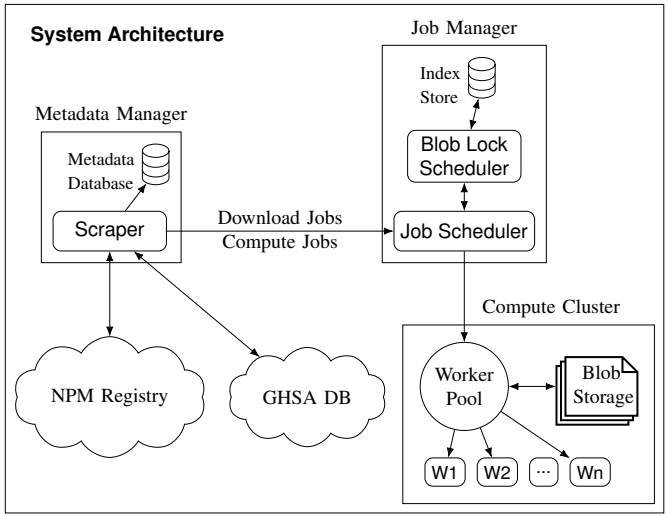
\begin{figure}
  \centering
  \resizebox{\columnwidth}{!}{%
    \begin{tikzpicture}[
        show background rectangle]

      \node[database, label={[align=left]\small Metadata\\\small Database}, scale=2] (MDB) {};
      \node[module] (SCRAPER) [below left=0.5cm and 0cm of MDB] {Scraper};
      \node[cloud, draw,cloud puffs=10,cloud puff arc=120, aspect=2, inner ysep=1em, below=1.5cm of SCRAPER] (REGISTRY) {NPM Registry};
      \node[cloud, draw,cloud puffs=10,cloud puff arc=120, aspect=2, inner ysep=1em, right=0.5cm of REGISTRY] (GHSA) {GHSA DB};


      \node[module] (JS) [right=4.0cm of SCRAPER] {Job Scheduler};
      \node[module, align=center] (BLS) [above=0.5cm of JS] {Blob Lock\\Scheduler};
      \node[database, label={[align=left]\small Index\\\small Store}, scale=2] (REDIS) [above right=0.6cm and -0.5cm of BLS] {};
      \node[circle, align=center, draw, minimum width=0.5cm, minimum height=0.5cm] (WP)  [below=1.6cm of JS]  {Worker\\Pool};
      \node[
        shape=document,
        double copy shadow={
            shadow xshift=-0.5ex,
            shadow yshift=-0.5ex
          },
        draw,
        fill=white,
        line width=1pt,
        text width=1cm,
        minimum height=1.0cm,
        align=center,
      ] (BLOB) [right=1cm of WP] {Blob\\Storage};
      \node[module, align=center, draw, minimum width=0.5cm, minimum height=0.5cm] (W2)  [below left=0.7cm and -1.5cm of WP] {\small W2};
      \node[module, align=center, draw, minimum width=0.5cm, minimum height=0.5cm] (W1)  [left=0.2cm of W2]  {\small W1};
      \node[module, align=center, draw, minimum width=0.5cm, minimum height=0.5cm] (WDOTS)  [right=0.2cm of W2]  {\small ...};
      \node[module, align=center, draw, minimum width=0.5cm, minimum height=0.5cm] (WN)  [right=0.2cm of WDOTS]  {\small Wn};

      \node[fit=(SCRAPER) (MDB), label=Metadata Manager, draw, inner sep=2mm] (fit1) {};
      \node[fit=(JS) (BLS) (REDIS), label=Job Manager, draw, inner sep=2mm] (fit2) {};
      \node[fit=(WP) (BLOB) (W1) (W2) (WDOTS) (WN), label={[xshift=0.4cm]Compute Cluster}, draw, inner sep=3mm] (fit3) {};

      \draw[->] (SCRAPER)--(MDB);

      \draw[->] (SCRAPER) to node[sloped, align=center]{Download Jobs\\
        Compute Jobs} (JS) ;

      \draw[->] (JS)--(WP);
      \draw[<->] (JS)--(BLS);
      \draw[<->] (BLS)--(REDIS);
      \draw[<->] (SCRAPER)--(REGISTRY);
      \draw[<->] (SCRAPER)--(GHSA);
      \draw[<->, shorten <=0pt, shorten >=5pt] (WP)--(BLOB);
      \draw[->] (WP)--(W1);
      \draw[->] (WP)--(W2);
      \draw[->] (WP)--(WN);

      \node[font=\sffamily\bfseries] (lab1) [above right=1.4cm and -0.3cm of fit1.north west] {System Architecture};

    \end{tikzpicture}
  }
  \caption{
    Overview of our system architecture.
  }
  \label{fig:system-architecture}
\end{figure}

In order to perform our methodology, we needed a system that could scrape and store all metadata and tarball data,
and allow us to perform analyses and experiments on both the metadata and tarball data.  This system needs to be able to run
on our academic Slurm-backed~\cite{slurm-website,slurm-paper} HPC cluster.
To solve this problem, we designed our own system, organized into 3 primary components (\cref{fig:system-architecture}):
\begin{enumerate}
  \item The Metadata Manager, which continually scrapes data from NPM and periodically from the GitHub Security Advisory Database;
  \item the Job Manager, which receives job requests (typically tarball download or compute jobs) from the Metadata Manager and then
        coordinates job execution and distributed file system locks; and
  \item the Compute Cluster, in which we can spawn worker nodes and access a networked file system.
\end{enumerate}

We now explain how we accomplish the primary tasks required by our methodology.

\subsection{Metadata Acquisition}
\label{sec:architecture-metadata}

NPM stores metadata in a CouchDB database. CouchDB is a document-oriented JSON database and is a good fit for NPM because it is schemaless and allows for arbitrary nesting of JSON objects, such as the \texttt{package.json} file.
For performing data analysis we find it to be a poor fit due to the extremely loose structure. There is almost no validation of the
\texttt{package.json} files in the CouchDB, making it difficult to
use for analyses without first cleaning the data.

The Metadata Manager (top left of \cref{fig:system-architecture}) continually receives metadata changes from NPM via their
changes API~\cite{npm-changes-github}, validates those changes, and inserts the data into PostgreSQL~\cite{postgres-website}.
Additionally, the Metadata Manager periodically scrapes the GitHub Security Advisory Database and
imports the security metadata into PostgreSQL as well.
RQ1 and RQ2 can be answered entirely via issuing PostgreSQL queries to the Metadata Manager.

When metadata changes are received that contain URLs to new package tarballs, the Metadata Manager enqueues a tarball download job to then
be handled by the Job Manager.

\subsection{Tarball Data Acquisition and Compute Cluster}
\label{sec:architecture-tarballs}

For scraping and storing package tarballs, we need to be able to
store tens of millions of tarballs, while allowing for both
concurrent writes to the storage since new tarballs are downloaded continually,
as well as concurrent reads from the storage when performing analyses.

The worker nodes within the Compute Cluster are connected via a networked file system.
One interesting approach would be to use a technology such as Hadoop~\cite{hadoop-website} on top of the networked file system
to accomplish this. However, we did not explore this approach out of concern of Hadoop's scalability with regards
to storing many small files~\cite{hadoop-small-files} (our use case). In addition, we are are unsure if Hadoop
can run correctly and efficiently on top of a networked file system.

Instead, we store tarball data in a custom-built blob storage system stored on the networked
file system (bottom right of \cref{fig:system-architecture}). The Job Manager (top right of \cref{fig:system-architecture})
controls access to the blob storage, keeping track of byte offsets and coordinating locks for writing,
while individual worker nodes in the Compute Cluster perform the networked disk I/O.

Tarballs are downloaded when the Job Manager receives a download job request from the Metadata Manager, at which point it
assigns the download job to a worker node. Similarly, the Metadata Manager
may also send compute job requests, which the Job Manager handles by distributing to many worker nodes and optionally allowing each to perform
lockless read-only operations from the blob storage.

This system allows us to continually scrape and store tens of millions of tarballs, and to efficiently retrieve them for computation
when answering RQ4. Additionally, while RQ3 does not read from the blob storage, it follows the same compute workflow.

\subsection{Time-Traveling Dependency Resolver}
\label{sec:time-travel-impl}

In order to carry out our experiments outlined in \cref{sec:methodology} for RQ3, we needed
to be able to observe how a package's dependencies would have been solved at arbitrary points in NPM's history.
We built a proxy server that can be used with vanilla NPM to enable time-travel dependency resolving.

NPM's command line tool enables the user to specify a custom package registry to use in place of \texttt{npmjs.com}.
To use our time-traveling resolver, we specify a registry base URL pointing to our proxy server
that includes in the URL the timestamp to time-travel to. The proxy server
then receives the timestamp and can then rewrite responses from \texttt{npmjs.com} to remove versions of packages
after the timestamp. Since this does not rely on the rest of our system, it is extremely easy
to setup and use. However, in order to scale the computation across the dataset,
we use the compute capabilities discussed above in \cref{sec:architecture-tarballs}.

\section{Results}
\label{sec:results}

At a high level, we would consider a package ecosystem to be healthy with regards to update distribution
when updates that are positive (performance improvements, bug fixes, security patches, etc.)
can be quickly and easily adopted by downstream dependencies, while disruptive changes
(security vulnerabilities, malware, etc.) flow more slowly.
In NPM, the flow of updates is determined by two factors:
how do downstream developers tend to specify version constraints for dependencies (RQ1),
and how do upstream developers tend to increment their version numbers when releasing updates (RQ2).
We start by explaining the overall structure and general properties of the dataset.
Then we move on to discuss RQ1 and RQ2 separately, and finally we consider how
RQ1 and RQ2 intersect in practice in the ecosystem (RQ3), and how they are related
to the actual contents of the updates (RQ4).

\subsection{Dataset Structure and General Properties}
\label{sec:dataset-structure}

As discussed in \cref{sec:intro} our collected data is split into two parts:
\begin{inparaenum}
  \item Ecosystem Metadata:
  This includes the full list of packages (2,663,681 packages), versions of every package (28,941,927 versions), and metadata for every version
  including version upload times, version numbers, dependencies, descriptions, links to repositories, and more.
  We also have a full scrape of all security advisories for NPM packages, including data on which versions are vulnerable
  and which version(s) patch the vulnerability.
  \item Tarballs of published packages:
  The full source tarball of every version of every package\footnote{excluding deleted content, which we describe in \cref{sec:threats-to-validity}} has been downloaded by our system.
\end{inparaenum}

\begin{figure*}
  \begin{center}
    \begin{subfigure}{0.3\textwidth}
      \includegraphics[width=\columnwidth]{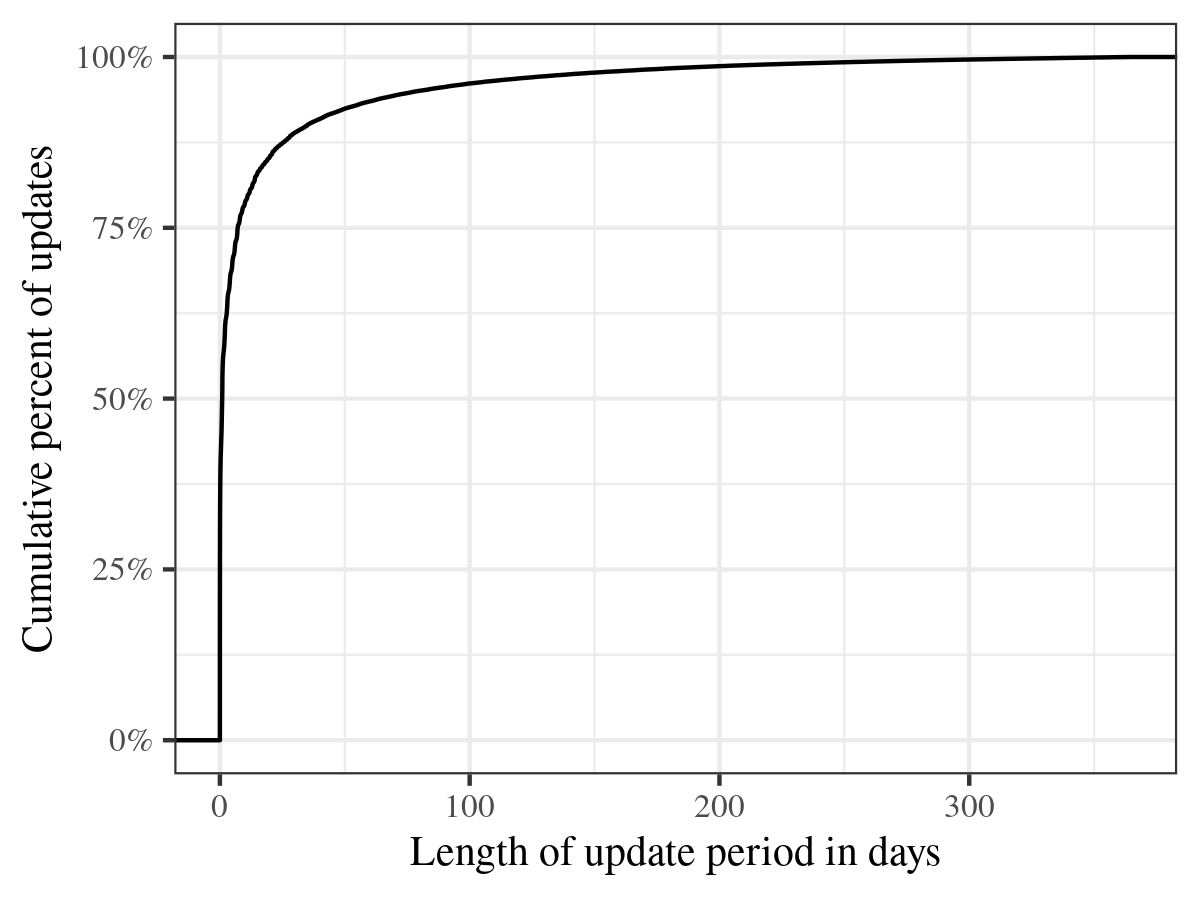}
      \caption{
        An ECDF of the time in days between the publication of two versions of a package.
        Note that this plot specifically excludes updates for non-prerelease versions
        of packages.
      }
      \label{plot:update-speed-ecdf}
    \end{subfigure}
    \quad
    \begin{subfigure}{0.3\textwidth}
      \includegraphics[width=\columnwidth]{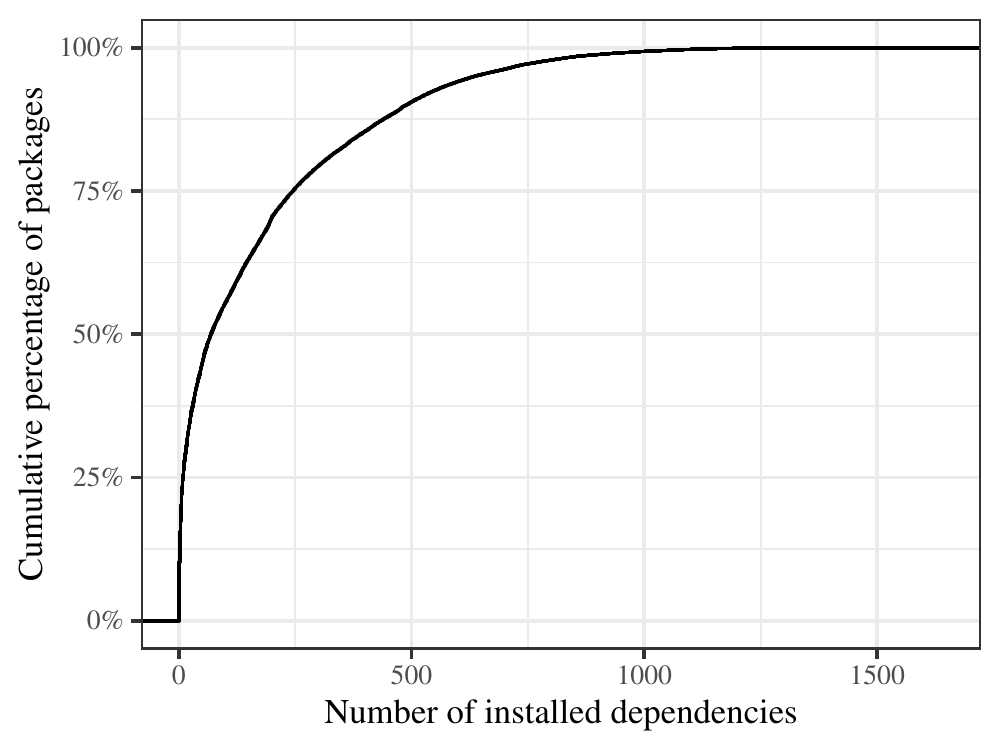}
      \caption{
        An ECDF of the number of (transitive) dependencies of each package. This was
        collected by resolving the latest version of every package on NPM
        as part of the experiment in \cref{plot:oldness-ecdfs}.
      }
      \label{plot:num-deps-ecdf}
    \end{subfigure}
    \quad
    \begin{subfigure}{0.3\textwidth}
      \includegraphics[width=\columnwidth]{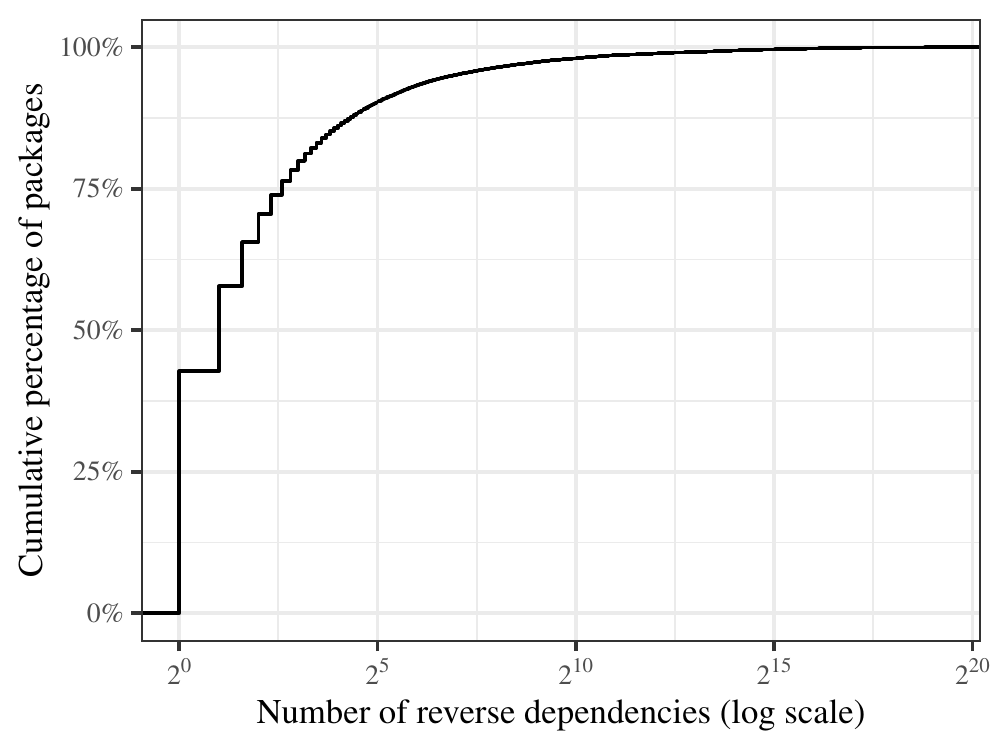}
      \caption{
        An ECDF of the number of reverse (transitive) dependencies of each package.
        Note that the x-axis is log-scaled.
        This was collected by resolving the latest version of every package on NPM
        as part of the experiment in \cref{plot:oldness-ecdfs}.
      }
      \label{plot:num-rev-deps-ecdf}
    \end{subfigure}
  \end{center}

  \caption{
    ECDF plots of general properties of the
    NPM ecosystem with regards to versioning and dependencies.
  }
  \label{plot:general-ecdfs}
\end{figure*}

Before diving into the core research questions, we first discuss general properties of the dataset.
\cref{plot:general-ecdfs} displays three distributions regarding our main objects of interest:
updates and dependencies.

\cref{plot:update-speed-ecdf} displays an ECDF (empirical cumulative distribution function) of the distribution of the time
between updates of packages, computed across 1,401,510 packages and 16,547,653 mined updates
(\cref{sec:methodology:package-updates}).
A surprising finding is how quickly updates are pushed out in many cases, with 25\%
of updates spanning only 39.87 \emph{minutes} or less, and 50\% of updates spanning
22.71 hours or less. However, a long tail of updates exists, with the top 25\% of updates spanning
7.78 days or longer, and 10\% spanning 40.12 days or longer. On average,
updates span 21.03 days.
A manual inspection of the data suggests that update behavior is quite bursty, with developers
releasing multiple updates in rapid succession, and then going silent for long periods of time;
however, this hypothesis should be investigated more thoroughly.

\cref{plot:num-deps-ecdf,plot:num-rev-deps-ecdf} display ECDFs of the distributions of
the numbers of (transitive) dependencies and downstream packages (i.e. transitive reverse dependencies),
respectively.
We selected the most recent non-prerelease version of every package with at least one update
(to filter out abandoned packages), yielding 1,401,510 packages.
We then used our time-traveling variant of NPM to resolve their dependencies and collect transitive dependency relations between packages,
disregarding versions.
Solving dependencies failed on some packages, due to both
true solving failures with NPM (e.g. missing dependencies) and transient system failures (discussed more in \cref{sec:threats-to-validity})
in the compute cluster.
In total, our experiments include successful executions of NPM's dependency solver on 696,419 packages.
The data shows that on average packages have 167.87 dependencies,
and 95\% of packages have solution sizes of
636 or fewer dependencies, with the largest solutions reaching up to 1,641 dependencies.

When turning to downstream packages however (\cref{plot:num-rev-deps-ecdf}),
the situation is quite asymmetrical, as there is a vastly longer tail of packages
with massive amounts of downstream packages. The top 3 depended-upon packages that we
observed were: \begin{inparaenum}
  \item \texttt{supports-color} (does a terminal support color?, 624,883 downstream packages),
  \item \texttt{debug} (logging library, 571,547 downstream packages), and
  \item \texttt{ms} (time conversion library, 515,684 downstream packages).
\end{inparaenum}
On the other hand, a large amount of packages are unused except by a handful of downstream packages, with 50\%
of packages having 2 or fewer downstream packages, and 90\% only being used by 30 or fewer
downstream packages.

\subsection{RQ1: Version Constraint Usage}

\begin{figure}
  \includegraphics[width=\columnwidth]{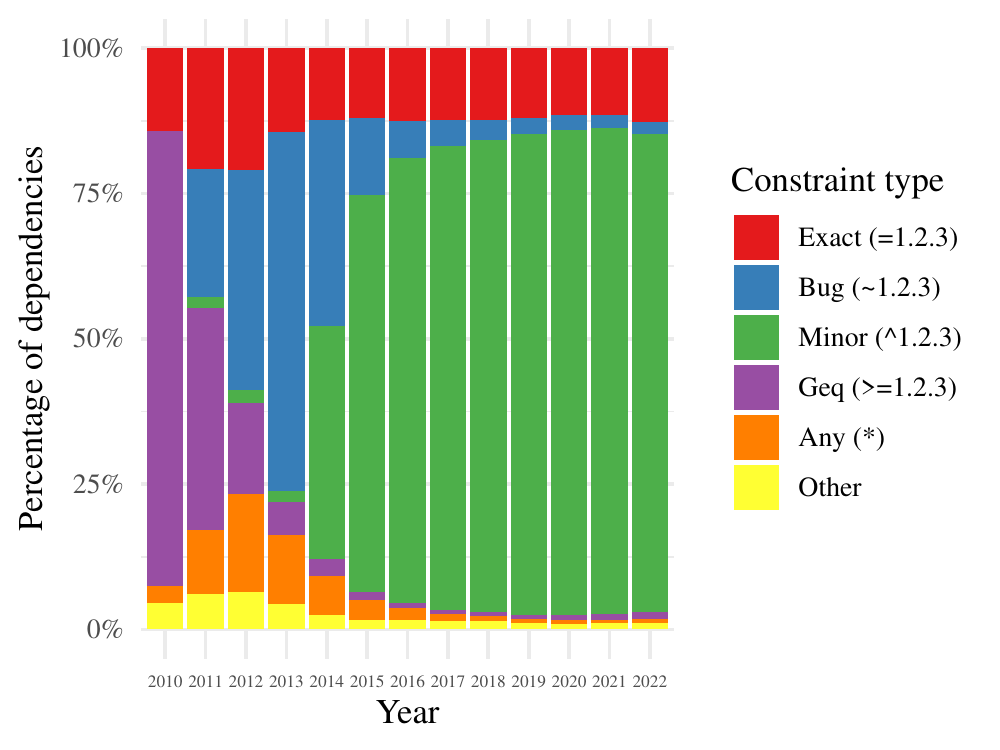}
  \caption{
    The relative popularity of each version constraint type
    across time. Percentages are \emph{not} cumulative over past years,
    reflecting only published dependencies within each year.
  }
  \label{plot:constraints-over-time}
\end{figure}

As described in \cref{sec:methodology:version-constraint-usage}, developers can specify version constraints in different ways, which controls the installation of newer versions of those dependencies.
\cref{plot:constraints-over-time} shows the frequency
of each main type of version constraint published in each year since 2010,
the year that NPM launched.
For each year, we include only packages that had at least one release published, and if a package released multiple versions in that year, we include only the most recently published non-prerelease
version that year.
In 2022, there were a total 429,265 packages with at least one release, and across all the years
1,678,681 distinct packages.

There are several interesting trends in constraint usage over time.
First, about 78.36\% of all initial dependencies were specified as
accepting any versions greater than some particular version (Geq, purple bars),
such as \texttt{"react" : ">= 1.2.3"}. Developers then
abandoned using Geq constraints within the first 3-4 years of NPM, likely because
they became unmaintainable as libraries began to introduce breaking changes that
would be automatically applied by Geq constraints.
Second, even though constraints that are flexible
in the minor component (Minor, green bars) currently represent a majority
of dependencies, the phenomenon of using minor flexible constraints
only started in 2014, and then rapidly expanded after. The expansion
of minor flexible constraints coincides with the decreased usage of
bug component flexible constraints (Bug, blue bars).
Third, developers have recently gravitated towards using only two types of constraints almost exclusively:
exact version constraints (Exact, red bars) and minor component flexible constraints.
Together, those types represent over 94.85\% of constraints in 2022.
Finally, the percentage of dependencies that are potentially able to automatically
receive updates (everything below the red bars) has stayed relatively
stable throughout the entire life of NPM, and is currently about 87.32\%
of all dependencies.


\subsection{RQ2: Semantic Versioning in Updates}
\label{section:results-rq2}


\begin{figure}
  \includegraphics[width=\columnwidth]{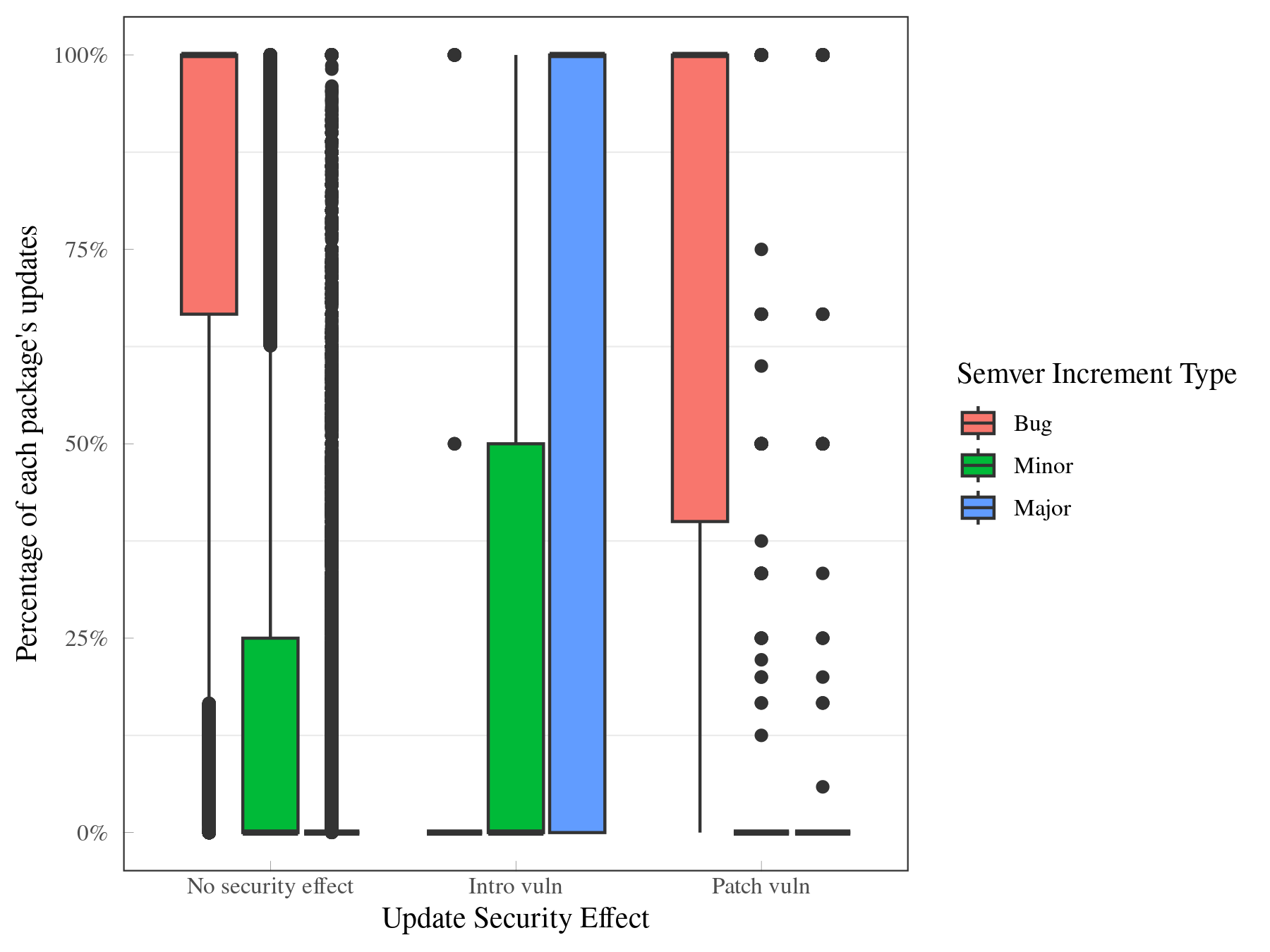}
  \caption{
    A boxplot visualizing the distribution of percentages of
    packages' updates by semver increment type, segmented
    across security effects. Within each security effect
    the percentages across semver increment types are normalized.
  }
  \label{plot:update-types-and-security}
\end{figure}
While RQ1 examined the usage of semantic versioning when specifying dependencies, RQ2 examines the usage of semantic versioning in deploying releases of those dependencies.
\cref{plot:update-types-and-security} displays boxplots
where each observation represents what percentage of a package's updates are one of the
three semver increment types, normalized across security effect.
This analysis includes 1,401,510 packages and 16,547,653 updates, as described in \cref{sec:methodology:package-updates}.

We find that in the no security effect category (the vast majority of updates),
the most common updates by far are bug semver increments, with 75\% of packages
having 66\% or more (lower quartile of left-most red box). Next most popular are
minor semver increments, and finally least most popular are major semver increments.

However, when we consider updates that introduce vulnerabilities, we see a different story.
Most packages introduce vulnerabilities via major semver increments,
indicating that vulnerabilities are often introduced when packages developers
release major new versions possibly consisting of many new features and significant
structural changes to the code base. We did however find 29 outlier packages that
introduced a vulnerability in at least one bug update.
A particularly interesting example is an update to the \texttt{ssri} package
(a cryptographic subresource integrity checking library, 23M weekly downloads)
from version 5.2.1 to 5.2.2. The update attempted to patch a regular expression denial of service vulnerability,
but inadvertently increased the severity of the vulnerability by changing the worst-case behavior from
quadratic to exponential complexity~\cite{ssri-vuln}.
This highlights the challenge package developers face in needing to quickly release patches to vulnerabilities, while needing to be
extremely careful when working on security-relevant code and releasing it through bug updates that will
be easily distributed to downstream packages.

Finally, in the case of vulnerabilities being patched, almost all patches are released
as bug semver increments, which means that the 87.32\% of non-exact constraints
shown in \cref{plot:constraints-over-time} would potentially be able to receive them automatically.
However, a handful of outlier packages have released vulnerability patches as
non-bug updates (we found 358 such updates across 298 packages). From manual inspection, it appears
that many of these updates include the fix for the security vulnerability mixed in with
many other changes, rather than the vulnerability fix being released independently.
For example, update 1.6.0 to 1.7.0 of the \texttt{xmlhttprequest} package (1.2M weekly downloads)
fixed a high-severity code injection vulnerability~\cite{xmlhttprequest-vuln}.
The security-relevant part of the update is only 1 line,
but 892 lines were modified in the update.
Without further investigation we do not know why some developers have chosen
to include security patches as part of larger updates rather than as standalone updates.

\subsection{RQ3: Out-of-Date Dependencies and Update Flows}
\label{sec:results:rq3}

\subsubsection{How out-of-date are packages' dependencies?}

Version constraints and semver update types work in tandem to
control the flow of updates to downstream packages, across
many chains of transitive dependencies.
Whether a downstream package receives up-to-date dependencies
depends not only on the constraints at the downstream package
and the type of semver increment at the reverse dependency,
but also on packages in the middle of a transitive dependency chain.

In this experiment, we select the latest version of every package with at least one update (1,401,510 packages).
We then use our time-traveling variant of NPM to solve the package's dependencies at the time the latest version was uploaded ($T_P$).
We then observe which of its installed dependencies are out-of-date,
where a dependency with version $V_D$ and upload time $T_D$
is out-of-date if another version $V_D'$ of the dependency has an upload
time $T_D'$ such that $T_D < T_D' < T_P$ and $V_D < V_D'$.
We then define the out-of-date time as $T_D' - T_D$ for the largest such $T_D'$.
After accounting for transient system failures, 696,419 packages were solved successfully.


\begin{figure*}
  \begin{center}
    \begin{subfigure}{0.45\textwidth}
      \includegraphics[width=\columnwidth]{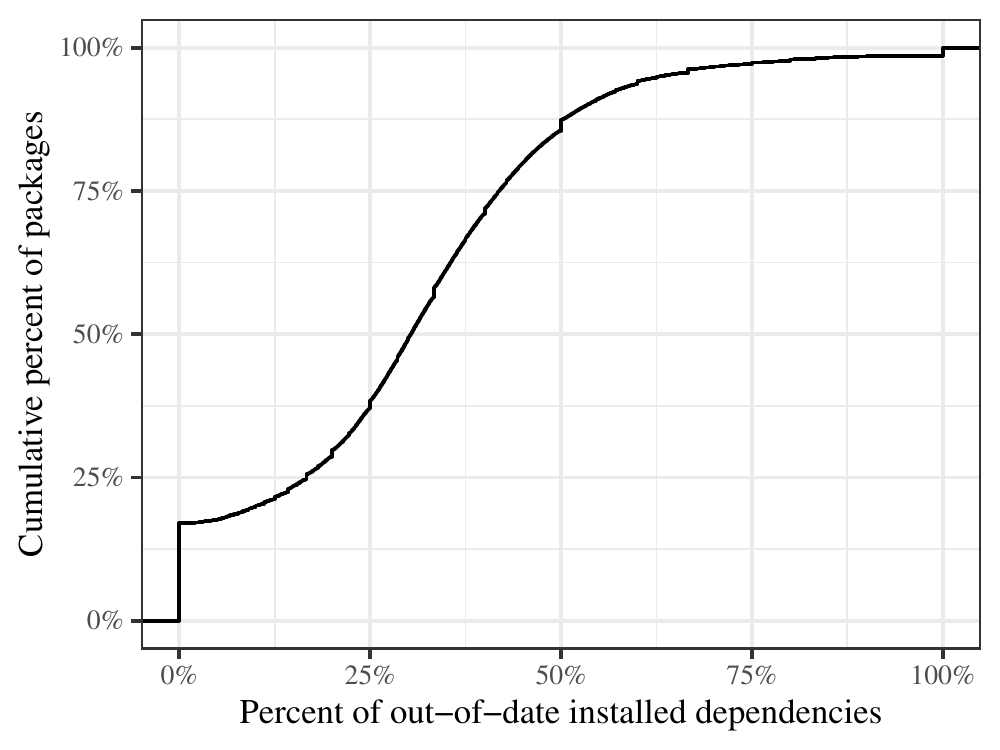}
      \caption{
        ECDF of percentage of each package's dependencies that are out-of-date.
      }
      \label{plot:oldness-perc-ecdf}
    \end{subfigure}
    \quad\quad
    \begin{subfigure}{0.45\textwidth}
      \includegraphics[width=\columnwidth]{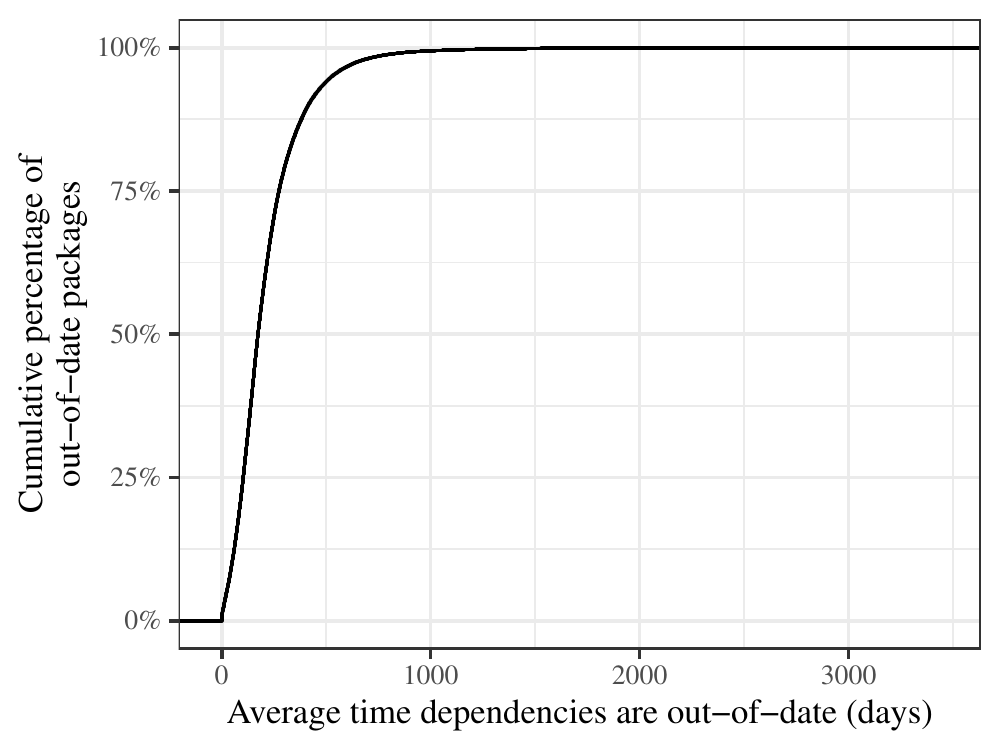}
      \caption{
        ECDF of average amount in days that dependencies are out-of-date by
        for each package with at least one-out-of-date dependency.
      }
      \label{plot:oldness-time-ecdf}
    \end{subfigure}
  \end{center}

  \caption{
    ECDF plots of technical lag distributions across the NPM ecosystem.
  }
  \label{plot:oldness-ecdfs}
\end{figure*}

\cref{plot:oldness-perc-ecdf} displays an ECDF of the distribution of
the percentage of each package's dependencies that are out-of-date.
There is a group of packages, about 17.08\%, that have fully up-to-date
dependencies. However, almost all of these have very few dependencies,
only 3.17 dependencies on average compared to 167.87 dependencies for the whole sample.
In other words, these fully up-to-date packages are packages that live primarily
on the far left side of the ECDF in \cref{plot:num-deps-ecdf}.

Moving beyond the spike of up-to-date packages,
most packages have at least some out-of-date dependencies, with
62.94\% of packages having 25\% or more of their
dependencies out-of-date.
Not only are packages often out-of-date, but they are often out-of-date for quite a while.
Among packages with at least one out-of-date
dependency, \cref{plot:oldness-time-ecdf} displays an ECDF of on average how out-of-date
each package's dependencies are. Half of all packages with out-of-date dependencies
have on average dependencies that are 173.87 days old or older, with a long tail
of 5\% of packages with dependencies that are on average 527.38 days old or older.
In contrast, updates are released within 21.03 days on average, and 50\%
are released within only 22.71 hours (\cref{plot:update-speed-ecdf}).

There can be a variety of reasons why packages have out-of-date dependencies,
some of which are intentional, such as developers choosing to stay on older versions
of libraries rather than rewrite code to handle breaking changes.

\subsubsection{How rapidly do updates flow downstream?}

We now wish to understand how updates flow to downstream packages,
and how developers respond when manual intervention is required.
For the most recent update prior to 2021 of every package, we randomly selected 50 downstream
packages that were up-to-date with the upstream package just prior to the update.
Using our time-traveling resolver we then solve the downstream package immediately after the update,
and in 1 day increments afterwards until the dependency on the old version of the
dependency has been updated or deleted.
In total, 888,294 update flows were successfully solved after accounting for transient system failures.

\begin{figure}
  \includegraphics[width=\columnwidth]{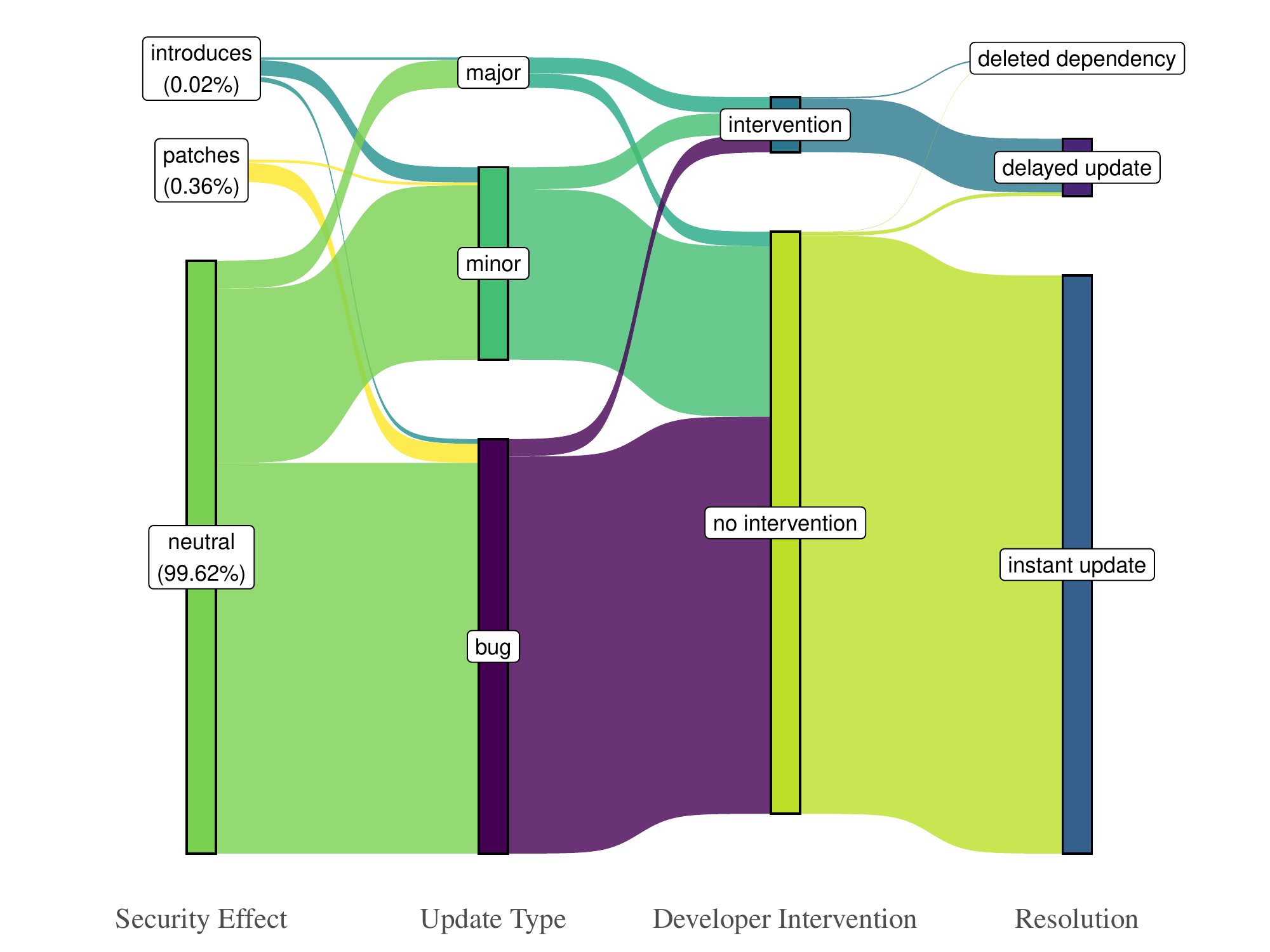}
  \caption{
    Visualization of update flow paths.
  }
  \label{plot:flow-analysis}
\end{figure}

\cref{plot:flow-analysis} visualizes the process of how updates flow
to downstream packages and how often developer intervention is required.
An update flow has multiple steps. First, the upstream (dependency) developer publishes the update
with a certain semver increment type (major, minor, or bug).
Once the update is marked as bug, minor, or major and uploaded to NPM,
it can then be received by downstream packages that depend on it, possibly
transitively. This can happen either automatically, by the downstream developer manually updating
or removing the dependency, or a developer in the middle of the transitive dependency chain
updating or removing the dependency.


Most commonly, downstream packages receive the update instantly and with no
human intervention needed ($\cdots \to \text{no intervention} \to \text{instant update}$).
This occurs when the package that declares
the constraint on the updated package uses a constraint that is at least
as flexible as the type of semver increment.
Note that the package declaring the constraint, and thus responsible for
allowing or inhibiting the update flow, could be either the final downstream package
or a package in the middle of the dependency chain.
This type of flow occurs for the majority of bug and minor updates,
which is induced by the distribution of constraint types (\cref{plot:constraints-over-time}).
As this type of flow is 90.09\% of all analyzed update flows, it is by far the
most common, indicating overall positive health among our random sample
of update flows through the NPM ecosystem.

The second most common update flow consists of updates that require
intervention from the developer of the downstream package
(and possibly developers of other packages as well),
and thus is delayed ($\cdots \to \text{intervention} \to \text{delayed update}$).
This occurs in 9.01\% of all analyzed update flows, and
involves a major update 28.11\% of the time,
a minor update 40.27\% of the time, and a bug update 31.62\% of the time.
Updates requiring intervention are due to constraints
that are more restrictive than the semver increment type.
Intervention thus involves developers either switching to a more flexible
constraint type or incrementing the constraint.

A small fraction (0.60\%) of updates are resolved not by the developer
of the downstream package performing an intervention, but by
a developer(s) in the middle ($\cdots \to \text{no intervention} \to \text{delayed update}$).
For this to occur, the developer of the package in the middle must have specified a
constraint that is too restrictive,
while the developer of the downstream package
specified a flexible enough constraint to allow for the intervention
of the package in the middle to be adopted. Since this type of flow
happens very rarely, this indicates that downstream packages typically have constraints that
are equally or more restrictive than their (transitive) dependencies.
This makes sense from a software engineering perspective, as the
deeper packages (those closer to libraries rather than applications)
have more incentive to use flexible constraints as they are likely to be reused
in contexts with otherwise conflicting constraints.

The final type of flow is when the out-of-date dependency is eventually deleted
rather than updated ($\cdots \to \text{intervention} \to \text{deleted dependency}$).
This occurs in only 0.29\% of all analyzed update flows,
indicating that developers do not generally delete dependencies.
More investigation could be done to understand why developers choose to delete
dependencies in a small number of cases.

\begin{figure}
  \includegraphics[width=\columnwidth]{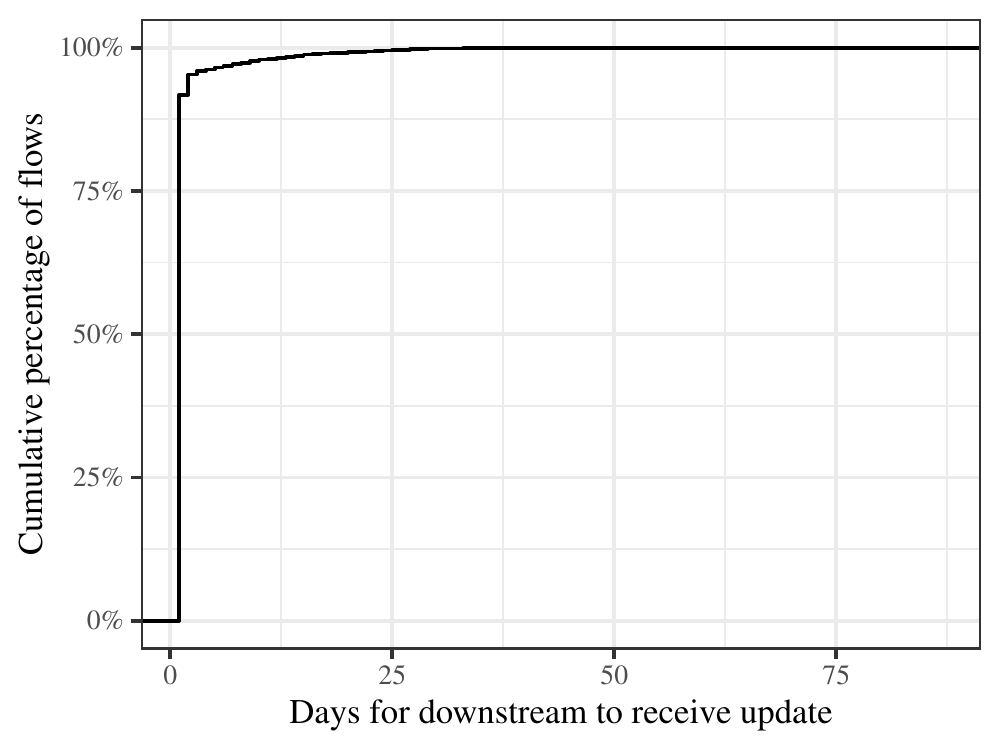}
  \caption{
    An ECDF plot of how long it takes for an update flow that is blocked
    to be resolved.
  }
  \label{plot:delayed-update-days}
\end{figure}

Among the update flows that are blocked due to restrictive constraints,
almost all update flows are unblocked via manual intervention quite rapidly.
\cref{plot:delayed-update-days} shows an ECDF of the distribution of how many
days it takes for each update flow to be unblocked.
The majority of blocked update flows (91.74\%) are unblocked within 1 day,
with a tail trailing off to 25 days or more. The surprising speed of
update flows being unblocked is due largely
to the fact that many packages that depend on each other are developed by the same contributors,
and they will often bump version numbers and update dependencies of their
packages nearly simultaneously.

Our results suggest that
most updates effectively flow to downstream packages,
while \cref{plot:oldness-ecdfs} suggests that most downstream packages have
at least some out-of-date dependencies. More investigation should
be carried out on this phenomenon, but we suspect this is
due in part to the number of dependencies per-package (\cref{plot:num-deps-ecdf}) and the rate
of updates (\cref{plot:update-speed-ecdf}). With packages having
an average of 167 dependencies, and updates being released on average every 21 days,
we would expect that for an average package every day multiple dependencies release updates and
potentially go out-of-date. Even with many updates being adopted instantly
or quickly, some dependencies will become stale.
This phenomena might also be explained by our methodology for this experiment, as we selected only packages that were already up-to-date at the time of our analysis.

\subsection{RQ4: Analyzing Code Changes in Updates}

We now turn to inspecting the \emph{contents} of package
updates rather than metadata analysis. Semantic versioning can only be
useful if package developers release updates that are in accordance
with what downstream packages expect from bug, minor, or major semver increments.
In this paper, we focus on providing a high-level characterization
of what updates generally consist of in the NPM ecosystem, across the different update types.
While more fine-grained analyses and related applications would be interesting and useful,
it is beyond the scope of this paper, and we defer discussion of ongoing and future work to \cref{sec:discussion}.
However, we believe that our dataset may be a useful building block
for evaluation within the active research area of update analysis systems.

\begin{figure}
  \includegraphics[width=\columnwidth]{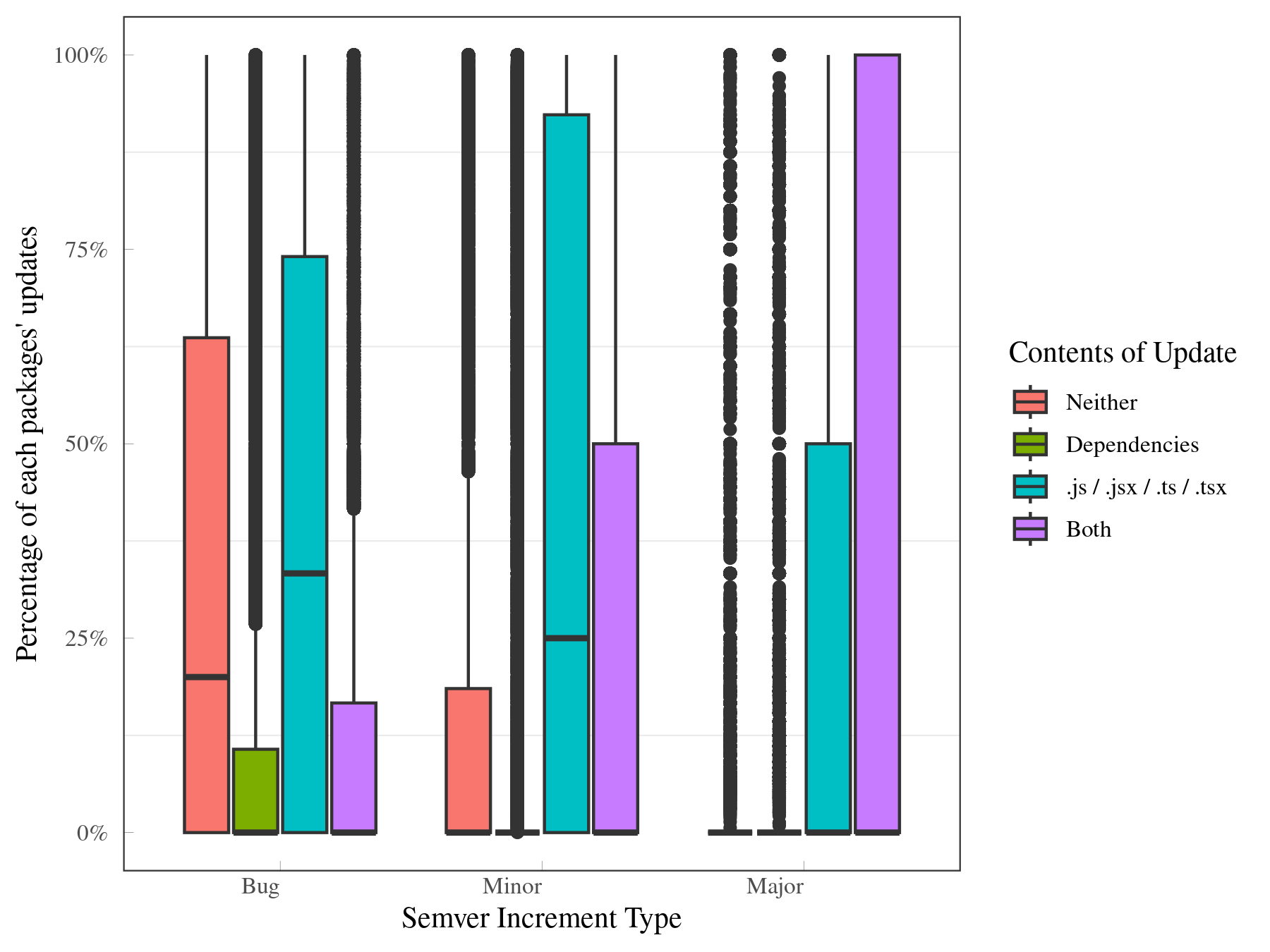}
  \caption{
    A boxplot displaying the distribution of the percentage of packages' updates grouped by
    semver increment type that change only code (\texttt{.js, .ts, .jsx, .tsx}), only dependencies,
    both, or neither.
  }
  \label{plot:contents-box}
\end{figure}

\cref{plot:contents-box} displays a boxplot where each observation is the
percentage of a package's updates within each semver increment type that
change only code (\texttt{.js}, \texttt{.ts}, \texttt{.jsx}, \texttt{.tsx}),
only dependencies, both, or neither. Note that updates categorized as neither may include
other changes such as modifications to other file types (README, CSS, etc.) or other metadata
changes besides dependencies. This uses the same set of packages and updates as from \cref{plot:update-types-and-security}, intersected with those we were able to successfully download tarballs
for, giving in total 1,339,684 packages and 14,903,021 updates.

First, we see that bug updates often contain no changes to code files, or to dependencies.
50\% of packages change neither code nor dependencies in about 20\%
or more of their bug updates, while 25\% of packages change neither in a majority
(64\%) of their bug updates.
A manual inspection of the data suggests that some of these updates consist of changes
to metadata (listed contributors, descriptions, READMEs) or to
configuration files (\texttt{.json}, \texttt{.yaml}, etc.), while other updates truly change nothing.
However, more investigation on our data could be done to quantify this more precisely.
Second, while it is not common to do so, 25\% of packages do occasionally release bug updates
which only modify dependencies (11\% or more of bug updates).
Looking at minor and major updates, the frequency of packages modifying neither or
only one or the other decreases, and when looking at major updates, most packages modify both
code and dependencies simultaneously.

\section{Discussion}
\label{sec:discussion}

Considering the results that we presented in \cref{sec:results}, we find a number of implications for software developers, ecosystem maintainers and researchers.
Developers consuming dependencies face persistent trade-offs between security, reliability, and technical lag.
We identify opportunities for ecosystem maintainers to reduce some of this friction
and point towards longer-term research directions to address some of the underlying challenges in package ecosystems.

\subsection{For Developers}

Our findings for RQ1 indicate that NPM has largely consolidated around using either exact or minor-flexible (\texttt{\^{}}) constraints, with the greatest proportion of dependencies specified as minor-flexible.
In practice this means that minor updates will flow to downstream packages nearly as easily as bug updates,
which we confirmed experimentally in RQ3.2, with 95.42\% of sampled bug updates and 86.55\% of sampled minor updates flowing automatically to downstream packages.
This finding is important for library maintainers, who might expect that downstream packages will manually inspect minor updates for compatibility.

Overall, there is a misalignment between the way that versions are released and the way that they are depended on,
as versions that are released as minor vs. bug updates commonly have distinct characteristics (\cref{plot:update-types-and-security,plot:contents-box}), while dependencies in downstream packages rarely distinguish between minor or bug updates (\cref{plot:constraints-over-time}).
Specifically, we find that 81.19\% of updates are released as bug updates, but 84.01\% of dependency constraints accept bug and minor updates.
While both bug and minor updates are supposed to maintain backwards compatibility, since minor updates may be more likely to include (inadvertent) breaking changes, developers may benefit in stability by using bug-flexible (\url{~}) constraints rather than minor-flexible constraints, which would still receive 81.19\% of updates.
This motivation may be even stronger for security-cautious developers as our results suggest that minor updates introduce vulnerabilities
more often than bug updates, however they must remain careful as even bug updates occasionally introduce vulnerabilities.

\subsection{For Ecosystem Maintainers}

Our findings in RQ2 indicated that some developers release security patches with minor and sometimes, even major version increments.
This finding is concerning as it makes it more difficult for downstream packages to receive the security fixes.
This suggests that ecosystems may benefit from ecosystem maintainers attempting to have tighter communication with
package developers around security patches, and help ensure that security patches are released in a timely manner,
with minimal changes, and as semver bug updates.

Our findings in RQ3.2 show a small fraction of update flows that are blocked by dependencies in the middle.
This is perhaps the most frustrating case for developers, as it is difficult to remedy the situation.
One option is to use NPM's overrides feature~\cite{npm-overrides}, which allows the downstream package to forcefully
override versions of transitive dependencies, even if this breaks version constraints.
While this can be effective in the short-term, one challenge is that the developer now has the maintenance burden of
removing the override when it is no longer necessary, or else face broken builds in the future.
To improve the developer experience, ecosystem maintainers could \begin{inparaenum}
  \item reduce the frequency of update propagation blockage by combining our analysis with centrality analysis to find critical packages that often block update flows, and work with them to address the situation; and
  \item improve ecosystem tooling around overrides to help developers automate the removal of overrides when no longer necessary.
\end{inparaenum}

\subsection{For Researchers}
\label{sec:discussion:researchers}

Our findings in RQ2 indicate that while NPM developers generally try to follow semver conventions,
they do not always do so consistently, and thus developers of downstream packages can not be entirely
confident about what exactly they will receive when updating dependencies (particularly if malicious developers release malware!).
This suggests a useful and broad design space of static or dynamic program analysis tooling that could help give
insight on what actually changes in an update. Such tools could aim to check for semver compliance~\cite{semantics-in-semantic-versioning,breakbot},
check that an update actually patches the claimed vulnerability correctly,
check for likely buggy changes in behavior~\cite{program-diffing},
or detect malware.
It may be particularly interesting to examine trends in semver compliance over time, as our analysis shows clear trends in the changing popularity of dependency constraints between 2010--2022.

There is already promising ongoing work in some of these directions, particularly malware detection~\cite{amalfi,weak-links-npm,snyk}
via metadata and lightweight syntactic features.
In RQ4 we found that a significant portion of packages publish bug updates that change neither
\texttt{js}, \texttt{ts}, \texttt{jsx}, \texttt{tsx} files, nor dependencies, which suggests
that a sizeable portion of updates may be changing other types of files,
and such changes may be an effective place for bad actors to hide malicious changes.
Whether the aim of this work is malware detection, bug detection, or other analyses, our results suggest
that such tooling should aim to handle
multiple file types, such as code, config files, embedded binaries, shell scripts, etc.

The analysis in RQ3.1 finds that there is a substantial amount of technical lag in NPM packages,
so tooling to help developers reduce technical lag could be quite impactful.
In our prior work we built a tool, {\sc MaxNPM}~\cite{maxnpm}, which allows developers to solve
dependencies in a way that minimizes technical lag (or other objectives)
while still satisfying current version constraints, but does not help when constraints themselves are out-of-date.
Complementary future research, such as what Jayasuriya suggests~\cite{towards-automated-updates}, could assist developers in performing these manual updates by helping with code migration
in response to breaking changes.

\section{Threats to Validity}
\label{sec:threats-to-validity}

\subsection{External Validity}

We were unable to reliably scrape packages that have been deleted (for malware, copyright violations, etc.)
or unpublished (voluntarily by the developer) from NPM, and thus we excluded these in our analyses.
For this reason our results might not generalize to malware or other types of packages that are often deleted.

Other than deleted packages, we consider the entire ecosystem, including
so called ``trivial'' packages~\cite{micro-packages,trivial-packages} and packages that
seem unimportant (e.g. few reverse dependencies).
We believe that it is difficult to tell if a package truly is irrelevant, as even a package with very few
reverse dependencies may in fact be an application that has been published on NPM. 
Furthermore, ``low-impact'' developers are nevertheless important as their experience with NPM matters for the future of the ecosystem.

We only obtain packages from NPM, and do not consider GitHub or other sources.
As such, this study may not generalize to JavaScript applications (rather than libraries),
as only some developers choose to publish their applications on NPM.
In addition, some developers may include dependencies by directly
copying source files into their packages, which we do not detect.
Finally, it is important to be careful when generalizing our results about security vulnerabilities,
as we are only able to obtain information about \emph{known} vulnerabilities,
which is likely a small subset of all vulnerabilities.

\subsection{Internal Validity}
Our system described in \cref{sec:system-architecture}
has a lot of moving parts,
and it is possible that there are bugs in our system that could
affect the results of our experiments. For example, we may have
missed some packages in our scraping process, or we may have
incorrectly downloaded some packages. We believe that this is
unlikely, as we have written unit tests for our system and have
tested it on a small subset of packages, and have not found any
bugs.

While running millions of package installations for RQ3 (\cref{sec:results:rq3})
we caused intermittent failures on our compute cluster by overflowing \texttt{/tmp}.
Since these failures were a function of system state and not of packages, we do not believe this
biased our results. To check this, we computed the mean and median of the number of
direct dependencies of successful and failed packages, and found that successful packages
have a mean of 9.91 direct dependencies and a median of 5, while the failed packages have a mean
10.93 direct dependencies and a median of 6. This suggests that failed packages were a bit larger,
but not enough to make our successful packages unrepresentative. Outliers with a large number of
dependencies existed with both failed and successful packages.

\subsection{Construct Validity}

Throughout RQ2--RQ4 we use our algorithm for computing updates as described in \cref{sec:methodology:package-updates}.
Since there is no ground truth for correctly mined updates, one may wish to consider refinements to this algorithm.
In particular, one may wish to have fine-grained equivalence classes by considering minor components as well.
However, this would not change the results where our algorithm already succeeds, and since the rejection rate is already quite low (1.8\%)
we did not believe a more complex algorithm justified the risk of analysis bugs.

In RQ4 we defined code changes to mean files with extensions
\texttt{.js}, \texttt{.ts}, \texttt{.jsx}, or \texttt{.tsx}.
This is because we wanted to focus on JavaScript and TypeScript code,
but this may have caused us to miss some JavaScript or TypeScript code with other extensions.
Depending on the purpose, future work might want to consider a broader
definition of what counts as code, such as shell scripts.

\section{Conclusion}
\label{sec:conclusion}

We present a large-scale analysis of semantic versioning in NPM, and a full, reusable dataset of
complete package metadata and tarball data from NPM. We find that there is a higher risk of
security vulnerabilities being introduced through minor rather than bug (i.e. patch) semver updates,
suggesting a motivation for developers to use bug-flexible constraints (\url{~}), even while
the NPM ecosystem has largely abandoned them in favor of minor-flexible constraints (\texttt{\^{}}).
While we find that most security patches are introduced in bug updates, we find a disturbing set of outliers that are released as minor or even major updates,
potentially causing slower adoption of security patches.
Future work examining the NPM ecosystem might build on our dataset and tooling, examining the contents of updates for bugs and/or vulnerabilities, along with mechanisms to mitigate technical lag.

\section{Data Availability}
\label{sec:data_avail}

Our artifact permanently archived on Zenodo~\cite{citeTheArtifact} contains our tools and the metadata from our dataset.
At \url{https://dependencies.science} we post:
\begin{inparaenum}
  \item continually updating snapshots of our data (including the contents of all packages), and
  \item the full implementations of both our scraping systems and our data analysis scripts.
\end{inparaenum}
We intend the site to be a useful resource for other researchers looking at NPM.

\section*{Acknowledgments}

We thank Northeastern Research Computing, especially Greg Shomo,
for computing resources and technical support.

  {
    \footnotesize

    \ifthenelse{\equal{\formattype}{\formattypeIEEE}} {
      \bibliographystyle{IEEEtran}
    } {
      \bibliographystyle{ACM-Reference-Format}
    }

    \bibliography{bib/venues-short,bib/main}
  }

\end{document}